\documentclass[12pt]{article}
\usepackage{amsmath}
\usepackage{bm}
\usepackage{color}

\begin{document}
\begin{center}
\begin{large}
{\bf Weak equivalence principle in noncommutative phase space and the parameters of noncommutativity}
\end{large}
\end{center}

\centerline {Kh. P. Gnatenko \footnote{E-Mail address: khrystyna.gnatenko@gmail.com}, V. M. Tkachuk \footnote{E-Mail address: voltkachuk@gmail.com}}
\medskip
\centerline {\small \it Ivan Franko National University of Lviv, Department for Theoretical Physics,}
\centerline {\small \it 12 Drahomanov St., Lviv, 79005, Ukraine}

\abstract{ The weak equivalence principle is studied in a space with noncommutativity of coordinates and noncommutativity of momenta. We find conditions on the parameters of noncommutativity which give the possibility to recover the equivalence principle in noncommutative phase space. It is also shown that in the case when these conditions are satisfied the motion of the center-of-mass of a composite system in noncommutative phase space and the relative motion are independent, the kinetic energy of composite system has additivity property and is independent on the systems composition. So, we propose conditions on the parameters of noncommutativity which give the possibility to solve the list of problems in noncommutative phase space.

}

\section{Introduction}

Noncommutativity has been recently received much attention owing to development of String Theory and Quantum Gravity (see, for example, \cite{Witten,Doplicher}). The idea that space might has a noncommutative structure was proposed by Heisenberg.  The scientist told his idea to Peierls, who informed Pauli about it, and Pauli told the idea to Oppenheimer \cite{Jackiw}. The first paper on the subject was written by the Oppenheimer's student, Snyder \cite{Snyder}.

Noncommutative phase space is characterized by the following relations for the coordinates and momenta
\begin{eqnarray}
[X_{i},X_{j}]=i\hbar\theta_{ij},\label{1alI00}\\{}
[X_{i},P_{j}]=i\hbar\delta_{ij},\\{}
[P_{i},P_{j}]=i\hbar\eta_{ij},\label{1alI01}
\end{eqnarray}
where $\theta_{ij}$, $\eta_{ij}$ are elements of constant antisymmetric matrices, parameters of noncommutativity.
In the classical limit $\hbar\rightarrow0$  the  quantum-mechanicalcommutator
is replaced by the Poisson bracket for corresponding classical variables.
From (\ref{1alI00})-(\ref{1alI01}) one obtains
\begin{eqnarray}
\{X_{i},X_{j}\}=\theta_{ij},\label{p1alI00}\\{}
\{X_{i},P_{j}\}=\delta_{ij},\\{}
\{P_{i},P_{j}\}=\eta_{ij}.\label{p1alI01}
\end{eqnarray}

Many papers are devoted to studies of physical systems in noncommutative phase space. Among them  hydrogen atom  \cite{Djemai,Kang,
Alavi,Bertolami}, harmonic oscillator \cite{Smailagic,Smailagic1,Hatzinikitas,Djemai,Li,Acatrinei,Giri,Geloun}, gravitational quantum well \cite{Bertolami1,Bastos} have been considered.

It is worth noting that noncommutativity  causes fundamental problems, among them the problem of violation of the weak equivalence principle \cite{Bastos1}. The weak equivalence principle (also called the uniqueness of free fall principle or the Galilean equivalence principle) states that kinematic characteristics, such as velocity and position of a point mass in a gravitational field depend only on its initial position and velocity, and are independent of its mass, composition and structure. This principle is a restatement of the equality of gravitational and inertial mass.

 In a space with noncommutativity of coordinates  the equivalence principle was considered in  \cite{Gnatenko1,Gnatenko3,Saha,Gnatenko2}. In paper \cite{Gnatenko1} we found condition on the parameter of noncommutativity for the recovering of the weak equivalence principle in a space with noncommutativity of coordinates. Namely, it was shown that in the case when the parameter of noncommutativity $\theta$, corresponding to the motion of a particle in noncommutative space, depends on its mass $m$ as $\theta\sim1/m$, the velocity and position of a body (composite system) in gravitation field do not depend on its mass and composition, so the weak equivalence principle is not violated.

In a space with noncommutativity of coordinates and noncommutativity of momenta (noncommutative phase space)  the equivalence principle was studied in  \cite{Bastos1,Bertolami2}. In \cite{Bertolami2} the authors concluded that the equivalence principle holds in noncommutative phase space in the sense that an accelerated frame of reference is locally equivalent to
a gravitational field, unless noncommutative parameters are anisotropic ($\eta_{xy}\neq$$\eta_{xz}$).

In the present paper we propose the conditions on the parameters of noncommutativity to recover the weak equivalence principle in a space with noncommutativity of coordinates and noncommutativity of momenta. We consider the general case when the coordinates and momenta of different particles satisfy noncommutative algebra  with different parameters of noncommutativity. It is shown that in two-dimensional noncommutative phase space
 \begin{eqnarray}
[X_{1},X_{2}]=i\hbar\theta,\label{alI0}\\{}
[X_{i},P_{j}]=i\hbar\delta_{ij},\\{}
[P_{1},P_{2}]=i\hbar\eta,\label{alI1}
\end{eqnarray}
here $\theta$, $\eta$ are constants, $i,j=(1,2)$, the position and the velocity of free-falling particle depend on its mass, therefore the weak equivalence principle is violated.
 We show that the equivalence principle is recovered in noncommutative phase space in the case when the parameters of noncommutativity, corresponding to the particle, are determined by its mass. Moreover, we conclude that the same conditions  give the possibility to obtain another important results. Among them are  preserving of the additivity property of the kinetic energy of composite system in noncommutative phase space; independence of the kinetic energy on the systems composition; independence of motion of the center-of-mass of composite system and relative motion.

 The article is organized as follows. In Section 2 we study the motion of a particle in gravitational field in noncommutative phase space and propose  conditions on the parameters of noncommutativity which give the possibility to recover the equivalence principle. In Section 3 the motion of composite system  is studied in gravitational field in noncommutative phase space and the equivalence principle is considered.  In Section 4 we examine the properties of kinetic energy of composite system and show that the same conditions, which give the possibility to recover the equivalence principle in noncommutative phase space, are important for preserving of the additivity of kinetic energy and its independence of composition. Conclusions are presented in Section 5.

\section{Weak equivalence principle in noncommutative phase space}

Let us first consider the classical motion of a particle in a uniform  gravitational field in two-dimensional noncommutative phase space and study the weak equivalence principle.

 In the classical limit, taking into account (\ref{alI0})-(\ref{alI1}), we have
 \begin{eqnarray}
\{X_{1},X_{2}\}=\theta,\label{palI0}\\{}
\{X_{i},P_{j}\}=\delta_{ij},\\{}
\{P_{1},P_{2}\}=\eta.\label{palI1}
\end{eqnarray}
In the case when uniform  gravitational field is directed along the $X_1$ axis, hamiltonian of  a particle of mass $m$  reads
\begin{eqnarray}
 H=\frac{P_{1}^{2}}{2m}+\frac{P_{2}^{2}}{2m}-mgX_1.
 \end{eqnarray}
 Taking into account (\ref{palI0})-(\ref{palI1}), we can write the following equations of motion
 \begin{eqnarray}
\dot{X}_1=\{X_1,H\}=\frac{P_{1}}{m},\label{form0220}\\
\dot{X}_2=\{X_2,H\}=\frac{P_{2}}{m}+mg\theta,\\
\dot{P_{1}}=\{P_1,H\}=mg+\eta\frac{P_2}{m},\\
\dot{P_{2}}=\{P_2,H\}=-\eta\frac{P_1}{m}.\label{form0221}
\end{eqnarray}
 From (\ref{form0220})-(\ref{form0221}), the trajectory of the particle in uniform gravitational field in noncommutative phase space reads
 \begin{eqnarray}
X_{1}(t)=\frac{A}{\eta}\sin\frac{\eta}{m}t-\frac{B}{\eta}\cos\frac{\eta}{m}t+C,\label{form0020}\\
X_{2}(t)=\frac{A}{\eta}\cos\frac{\eta}{m}t+\frac{B}{\eta}\sin\frac{\eta}{m}t-\frac{mg}{\eta}t+mg\theta t+D,\label{form0021}
\end{eqnarray}
here $A$, $B$, $C$, $D$ are constants. Considering the following initial conditions
 \begin{eqnarray}
 X_1(0)=X_{01},\label{form030}\\
 X_2(0)=X_{02},\\
\dot{X}_1(0)=\upsilon_{01},\\
\dot{X}_2(0)=\upsilon_{02},\label{form031}
\end{eqnarray}
we have
 \begin{eqnarray}
X_{1}(t)=\frac{m\upsilon_{01}}{\eta}\sin\frac{\eta}{m}t+\left(\frac{m^2g}{\eta^2}-\frac{m^2g\theta}{\eta}+\frac{m\upsilon_{02}}{\eta}\right)\left(1-\cos\frac{\eta}{m}t\right)+X_{01},\label{form0023}\\
X_{2}(t)=\left(\frac{m^2g}{\eta^2}-\frac{m^2g\theta}{\eta}+\frac{m\upsilon_{02}}{\eta}\right)\sin\frac{\eta}{m}t-\frac{m\upsilon_{01}}{\eta}\left(1-\cos\frac{\eta}{m}t\right)-\nonumber\\-\frac{mg}{\eta}t+mg\theta t+X_{02}.\label{form0024}
\end{eqnarray}
Note that in the limit $\eta\rightarrow0$ we obtain well known result
which corresponds to the ordinary space
 \begin{eqnarray}
X_{1}(t)=\frac{gt^2}{2}+\upsilon_{01}t+X_{01},\\
X_{2}(t)=\upsilon_{02}t+X_{02},
\end{eqnarray}
but the relation between momentum and velocity is changed, namely
\begin{eqnarray}
P_1=m\dot{X}_1,\\
P_2=m(\dot{X}_2+mg\theta).\label{form0}
\end{eqnarray}

Let us stress that according to (\ref{form0023})-(\ref{form0024}) the trajectory of a particle in  uniform gravitational field depends on its mass. So, the weak equivalence principle is violated in noncommutative phase space.
 Note that we face the problem of violation of the equivalence principle if we assume that parameters of noncommutativity $\eta$ and $\theta$ are the same for particles of different masses. In general case different particles may feel noncommutativity with different parameters.

 According to (\ref{form0023})-(\ref{form0024}) the trajectory of a particle in gravitational field depends on the combinations $m\theta$ and $\eta/m$.
It is important to mention that if the following conditions are satisfied
 \begin{eqnarray}
\theta m=\gamma=const,\label{ondN}\\
\frac{\eta}{m}=\alpha=const,\label{ondN2}
\end{eqnarray}
where $\gamma$ and $\alpha$ are constants which are the same for  particles of different masses, we have
 \begin{eqnarray}
X_{1}(t)=\frac{\upsilon_{01}}{\alpha}\sin \alpha t+\left(\frac{g}{\alpha^2}-\frac{g\gamma}{\alpha}+\frac{\upsilon_{02}}{\alpha}\right)\left(1-\cos \alpha t\right)+X_{01},\label{form0025}\\
X_{2}(t)=\left(\frac{g}{\alpha^2}-\frac{g\gamma}{\alpha}+\frac{\upsilon_{02}}{\alpha}\right)\sin \alpha t-\frac{\upsilon_{01}}{\alpha}\left(1-\cos \alpha t\right)-\nonumber\\-\frac{g}{\alpha}t+\gamma g t+X_{02}.\label{form0026}
\end{eqnarray}
 So, in the case when the parameters of noncommutativity $\theta_a$, $\eta_a$, corresponding to a particle, are determined by its mass $m_a$ as
 \begin{eqnarray}
\theta_a=\frac{\gamma}{m_a},\label{condN}\\
\eta_a=\alpha m_a,\label{condN2}
\end{eqnarray}
trajectory of particle (\ref{form0025})-(\ref{form0026}) does not depend on the mass. So,  conditions (\ref{condN}), (\ref{condN2}) lead to the  recovering of the weak equivalence principle in noncommutative phase space. Note also that when condition (\ref{condN2}) is satisfied according to (\ref{form0}) we obtain that momentum is proportional to the mass, $P_2=m(\dot{X}_2+\gamma g)$, as it has to be.

Let us consider more general case of motion of a particle in non-uniform gravitational field  in noncommutative phase space and study the equivalence principle. The hamiltonian of a particle of mass $m$ in gravitational field  $V(X_1,X_2)$ reads
 \begin{eqnarray}
 H=\frac{P_{1}^{2}}{2m}+\frac{P_{2}^{2}}{2m}+mV(X_1,X_2),
 \end{eqnarray}
where coordinates and momenta satisfy commutation relations (\ref{palI0})-(\ref{palI1}).
Taking into account (\ref{palI0})-(\ref{palI1}), we can write
 \begin{eqnarray}
\dot{X}_1=\{X_1,H\}=\frac{P_{1}}{m}+m\theta\frac{\partial V(X_1,X_2)}{\partial X_2},\label{form1020}\\
\dot{X}_2=\{X_2,H\}=\frac{P_{2}}{m}-m\theta\frac{\partial V(X_1,X_2)}{\partial X_1},\\
\dot{P_{1}}=\{P_1,H\}=-m\frac{\partial V(X_1,X_2)}{\partial X_1}+\eta\frac{P_2}{m},\\
\dot{P_{2}}=\{P_2,H\}=-m\frac{\partial V(X_1,X_2)}{\partial X_2}-\eta\frac{P_1}{m}.\label{form1021}
\end{eqnarray}

If we suppose that  parameters of noncommutativity $\theta$, $\eta$ are the same for different particles with different masses, from equations (\ref{form1020})-(\ref{form1021}) we can conclude that velocity of a particle in gravitational field  depends on its mass. Therefore, the equivalence principle is violated. The way to solve the problem of violation of the equivalence principle is to consider parameters of noncommutativity depending on the particles mass. In the case when conditions (\ref{condN}), (\ref{condN2}) are satisfied
we have
 \begin{eqnarray}
\dot{X}_1=\frac{P_{1}}{m}+\gamma\frac{\partial V(X_1,X_2)}{\partial X_2},\label{form20}\\
\dot{X}_2=\frac{P_{2}}{m}-\gamma\frac{\partial V(X_1,X_2)}{\partial X_1},\\
\dot{P_{1}}=-m\frac{\partial V(X_1,X_2)}{\partial X_1}+\alpha P_2,\\
\dot{P_{2}}=-m\frac{\partial V(X_1,X_2)}{\partial X_2}-\alpha P_1.\label{form21}
\end{eqnarray}
Equations (\ref{form20})-(\ref{form21}) can be rewritten as
 \begin{eqnarray}
\dot{X}_1=P^{\prime}_{1}+\gamma\frac{\partial V(X_1,X_2)}{\partial X_2},\label{formp20}\\
\dot{X}_2=P^{\prime}_{2}-\gamma\frac{\partial V(X_1,X_2)}{\partial X_1},\\
\dot{P^{\prime}_{1}}=-\frac{\partial V(X_1,X_2)}{\partial X_1}+\alpha P^{\prime}_2,\\
\dot{P^{\prime}_{2}}=-\frac{\partial V(X_1,X_2)}{\partial X_2}-\alpha P^{\prime}_1.\label{formp21}
\end{eqnarray}
here we use the following notation
 \begin{eqnarray}
P^{\prime}_{i}=\frac{P_{i}}{m}.
\end{eqnarray}
Note, that equations (\ref{formp20})-(\ref{formp21}) written in  terms of variables $X_i$, $P^{\prime}_i$  do not contain mass. So, their solution $X_i=X_i(t)$, $P^{\prime}_i=P^{\prime}_i(t)$ does not depend on the mass too.
Therefore we can conclude that  in the case when conditions (\ref{condN}), (\ref{condN2}) hold the kinematic characteristics of the particle depend on the parameters $\gamma$ and $\alpha$ and do not depend on its mass. So, the weak equivalence principle is recovered in noncommutative phase space.

In more general case it is worth to consider the motion of composite system in gravitational field in noncommutative phase space and examine the equivalence principle. For this purpose the detailed studies of motion of a composite system in noncommutative phase space are needed.

\section{Composite system in noncommutative phase space and the equivalence principle}

Let us study a system of two particles of masses $m_1$ and $m_2$ in two-dimensional noncommutative phase space.  In general case different particles may feel noncommutativity with different parameters. So, there is a problem of describing the motion of the center-of-mass of composite system made of different particles in noncommutative phase space.

We consider the following commutation relations
\begin{eqnarray}
[X_{1}^{(a)},X_{2}^{(b)}]=i\hbar\delta^{ab}\theta_{a},\label{al0}\\{}
[X_{i}^{(a)},P_{j}^{(b)}]=i\hbar\delta^{ab}\delta_{ij},\\{}
[P_{1}^{(a)},P_{2}^{(b)}]=i\hbar\delta^{ab}\eta_{a},\label{al1}
\end{eqnarray}
here $i=(1,2)$, $j=(1,2)$, indices $a$, $b$ label the particles, $\theta_{a}$, $\eta_{a}$ are parameters of noncommutativity, which correspond to a particle of mass $m_a$. The corresponding Poisson brackets read
\begin{eqnarray}
\{X_{1}^{(a)},X_{2}^{(b)}\}=\delta^{ab}\theta_{a},\label{pal0}\\{}
\{X_{i}^{(a)},P_{j}^{(b)}\}=\delta^{ab}\delta_{ij},\\{}
\{P_{1}^{(a)},P_{2}^{(b)}\}=\delta^{ab}\eta_{a},\label{pal1}
\end{eqnarray}
Let us study the following hamiltonian
\begin{eqnarray}
 H=\frac{({\bf P}^{(1)})^{2}}{2m_{1}}+\frac{({\bf P}^{(2)})^{2}}{2m_{2}}+U(|{\bf X}^{(1)}-{\bf X}^{(2)}|),\label{h}
 \end{eqnarray}
where $U(|{\bf X}^{(1)}-{\bf X}^{(2)}|)$ is the interaction potential energy. Introducing total momentum, coordinates of the center-of-mass, coordinates and momenta of relative motion in the traditional way
\begin{eqnarray}
\tilde{{\bf P}}={\bf P}^{(1)}+{\bf P}^{(2)},\label{form4}\\
\tilde{{\bf X}}=\frac{m_{1}{\bf X}^{(1)}+m_{2}{\bf X}^{(2)}}{m_{1}+m_{2}},\label{form04}\\
\Delta{\bf P}=\mu_{1}{\bf P}^{(2)}-\mu_{2}{\bf P}^{(1)},\label{form5}\\
{\Delta\bf X}={\bf X}^{(2)}-{\bf X}^{(1)},\label{form05}
\end{eqnarray}
 hamiltonian (\ref{h}) can be rewritten as follows
\begin{eqnarray}
H=\frac{(\tilde{{\bf P}})^{2}}{2M}+\frac{(\Delta{\bf P})^{2}}{2\mu}+U(|\Delta{\bf X}|),
\end{eqnarray}
here $M=m_{1}+m_{2}$ is the total mass and $\mu=m_{1}m_{2}/(m_{1}+m_{2})$ is the reduced mass.

It is important to mention that in noncommutative phase space the Poisson bracket for total momentum and the momentum of relative motion is not equal to zero
 \begin{eqnarray}
\{\tilde{P}_1,\Delta{P}_2\}=-\{\tilde{P}_2,\Delta{P}_1\}=\mu_1\eta_2-\mu_2\eta_1.
\end{eqnarray}
Note also that the coordinates of the center-of-mass and the coordinates of relative motion satisfy the following relation
\begin{eqnarray}
\{\tilde{X}_{1},\Delta X_{2}\}=\frac{m_{2}\theta_{2}-m_{1}\theta_{1}}{m_{2}+m_{1}}.
\end{eqnarray}
So, the motion of the center-of-mass and the relative motion are not independent in noncommutative phase space. In noncommutative phase space  two-particle problem can not be reduced to a one-particle problem for the internal motion.

Let us note here that the Poisson bracket for total momentum and the momentum of relative motion is equal  to zero,
$\{\tilde{P}_1,\Delta{P}_2\}=\{\tilde{P}_2,\Delta{P}_1\}=0$,
in the case when the following condition on the parameters of momentum noncommutativity is satisfied
\begin{eqnarray}
\frac{\eta_1}{m_1}=\frac{\eta_2}{m_2}=\alpha=const,\label{cond}
\end{eqnarray}
where $\alpha$ is a dimensionless constant.

 It is worth also to note that Poisson bracket for the coordinates of the center-of-mass and the coordinates of relative motion vanishes in the case when the parameters of coordinate noncommutativity satisfy the following condition
\begin{eqnarray}
\theta_1 m_1=\theta_2 m_2=\gamma=const,\label{cond2}
\end{eqnarray}
here $\theta_i$ is the parameter of noncommutativity which corresponds to a particle of mass $m_i$, and $\gamma$ is a constant.

So, the two-particle problem can  be reduced to a one-particle problem for the internal motion in noncommutative phase space if parameters of noncommutativity which corresponds to a particle are determined by its mass as (\ref{cond}), (\ref{cond2}). It is important to stress that conditions  (\ref{cond}), (\ref{cond2}) are the same as conditions (\ref{condN}), (\ref{condN2}) proposed for the recovering of the weak equivalence principle.

Let us consider Poisson brackets for the total momenta and the coordinates of the center-of-mass. Taking into account (\ref{pal0})-(\ref{pal1}), (\ref{form4}), (\ref{form04}) we can write the following  relations
\begin{eqnarray}
\{\tilde{X}_1,\tilde{X}_2\}=\tilde{\theta},\\{}
\{\tilde{P}_1,\tilde{P}_2\}=\tilde{\eta},\\{}
\{\tilde{X}_i,\tilde{P}_j\}=\delta_{ij},{}
\end{eqnarray}
where $\tilde{\theta}$ and $\tilde{\eta}$ are an effective parameters of noncommutativity which are defined as
\begin{eqnarray}
\tilde{\theta}=\frac{m_{1}^{2}\theta_{1}+m_{2}^{2}\theta_{2}}{(m_{1}+m_{2})^{2}},\label{0}\\
\tilde{\eta}=\eta_1+\eta_2.\label{01}
\end{eqnarray}
Also, from (\ref{pal0})-(\ref{pal1}), (\ref{form5}), (\ref{form05}) we can write the following relations for coordinates and momenta of relative motion
\begin{eqnarray}
\{\Delta{X}_1,\Delta{X}_2\}=\Delta\theta,\\{}
\{\Delta{P}_1,\Delta{P}_2\}=\Delta\eta,\\{}
\{\Delta{X}_i,\Delta{P}_j\}=\delta_{ij},{}
\end{eqnarray}
here parameters $\Delta\theta$ and $\Delta\eta$ read
\begin{eqnarray}
\Delta{\theta}=\theta_{1}+\theta_{2},\label{02}\\
\Delta{\eta}=\mu_2^2\eta_1+\mu_1^2\eta_2.\label{03}
\label{form7}
\end{eqnarray}

So, the coordinates of the center-of-mass and the total momenta satisfy noncommutative algebra with effective parameters of noncommutativity (\ref{0}), (\ref{01}), which depend on the masses of particles forming the system and on the parameters of noncommutativity $\theta_i$, $\eta_i$, corresponding to the individual particles.  Let us mention that in the case when conditions (\ref{cond}), (\ref{cond2}) are satisfied effective parameters $\tilde{\theta}$, $\tilde{\eta}$ depend only on the total mass of the system and do not depend on its composition. We have
\begin{eqnarray}
\tilde{\theta}=\frac{\gamma}{M},\label{ef}\\
\tilde{\eta}=\alpha M.\label{eff}
\end{eqnarray}

It is worth also mentioning that conditions  (\ref{cond}), (\ref{cond2}) are satisfied for effective parameters of noncommutativity $\tilde{\theta}$, $\tilde{\eta}$ and parameters $\Delta\theta$, $\Delta\eta$.  Taking into account  (\ref{cond}), (\ref{cond2}), from (\ref{0}), (\ref{01}), (\ref{02}), (\ref{03}) we have
\begin{eqnarray}
\frac{\tilde{\eta}}{M}=\frac{\Delta\eta}{\mu}=\frac{\eta_1}{m_1}=\frac{\eta_2}{m_2}=\alpha=const.\\
\tilde{\theta} M=\Delta\theta \mu=\theta_1 m_1=\theta_2 m_2=\gamma=const.
\end{eqnarray}

In more general case of system made of $N$ particles  we have the following hamiltonian
 \begin{eqnarray}
  H=\sum_{a}\frac{({\bf P}^{(a)})^{2}}{2m_{a}}+\frac{1}{2}\mathop{\sum_{a,b}}\limits_{a\neq b} U(|{\bf X}^{(a)}-{\bf X}^{(b)}|),
 \end{eqnarray}
here coordinates and momenta satisfy  (\ref{pal0})-(\ref{pal1}).   The total momenta, coordinates of the center-of-mass, coordinates and momenta of relative motion read
\begin{eqnarray}
\tilde{{\bf P}}=\sum_{a}{\bf P}^{(a)},\label{05}\\
\tilde{{\bf X}}=\sum_{a}\mu_{a}{\bf X}^{(a)},\label{00005}\\
\Delta{\bf P}^{{a}}={\bf P}^{(a)}-\mu_{a}\tilde{{\bf P}},\\
{\Delta\bf X}^{(a)}={\bf X}^{(a)}-\tilde{{\bf X}}.\label{06}
\end{eqnarray}
Taking into account (\ref{pal0})-(\ref{pal1}), we can write
\begin{eqnarray}
\{\tilde{X}_1,\tilde{X}_2\}=\tilde{\theta},\label{07}\\{}
\{\tilde{P}_1,\tilde{P}_2\}=\tilde{\eta},\\{}
\{\tilde{X}_i,\tilde{P}_j\}=\{\Delta{X}_i,\Delta{P}_j\}=\delta_{ij},\label{08}\\{}
\{\Delta{X}_1^{(a)},\Delta{X}_2^{(b)}\}=-\{\Delta{X}_2^{(a)},\Delta{X}_1^{(b)}\}=\delta^{ab}\theta_{a}-\mu_{a}\theta_{a}-\mu_{b}\theta_{b}+\tilde{\theta} ,\\{}
\{\Delta{P}_1^{(a)},\Delta{P}_2^{(b)}\}=-\{\Delta{P}_2^{(a)},\Delta{P}_1^{(b)}\}=\delta^{ab}\eta_a-\mu_b\eta_a-\mu_a\eta_b+\mu_a\mu_b\tilde{\eta},{}
\end{eqnarray}
with effective parameters of noncommutativity $\tilde{\theta}$, $\tilde{\eta}$ given by
\begin{eqnarray}
\tilde{\theta}=\frac{\sum_{a}m_{a}^{2}\theta_{a}}{(\sum_{b}m_{b})^{2}},\label{eff}\\
\tilde{\eta}=\sum_{a}\eta_a.\label{eff2}
\end{eqnarray}
Note that the following relations are satisfied
 \begin{eqnarray}
 \{\tilde{X}_{1},\Delta X_{2}^{(a)}\}=-\{\tilde{X}_{2},\Delta X_{1}^{(a)}\}=\mu_{a}\theta_{a}-\tilde{\theta},\\{}
 \{\tilde{P}_1,\Delta{P}^{a}_2\}=-\{\tilde{P}_2,\Delta{P}^{a}_1\}=\eta_a-\mu_a\sum_{b}\eta_b.
\end{eqnarray}
 It is important to stress that in the case when relations (\ref{condN}), (\ref{condN2}) hold
the Poisson brackets for the total momenta and the momenta of relative motion and  the Poisson brackets for coordinates of the center-of-mass and the coordinates of relative motion vanish $ \{\tilde{X}_{1},\Delta X_{2}^{(a)}\}=-\{\tilde{X}_{2},\Delta X_{1}^{(a)}\}=0$, $\{\tilde{P}_1,\Delta{P}^{a}_2\}=-\{\tilde{P}_2,\Delta{P}^{a}_1\}=0.$  In the case the motion of the center-of-mass and the relative motion are independent in noncommutative phase space.
In addition,  the effective parameters of noncommutativity (\ref{eff}), (\ref{eff2}) which describe the motion of the center-of-mass of the system do not depend on its composition if relations (\ref{condN}), (\ref{condN2}) hold. We have  $\tilde{\theta}=\gamma/M$ and $\tilde{\eta}=\alpha M$.

So, conditions (\ref{ondN}), (\ref{ondN2}) which are proposed for the recovering of the weak equivalence principle  are also important in consideration of composite system in noncommutative phase space.

Taking into account the results presented above, let us consider the motion of composite system (macroscopic body) in gravitational filed in noncommutative phase space and study the equivalence principle.
 For composite system in gravitational field  $V(\tilde{X}_1,\tilde{X}_2)$ we have the following hamiltonian
 \begin{eqnarray}
 H=\frac{\tilde{{\bf P}}^{2}}{2M}+M V(\tilde{X}_1,\tilde{X}_2)+H_{rel},
 \end{eqnarray}
where    $M$ is the total mass of the system,  $\tilde{X}_i$, $\tilde{P}_i$,  are  coordinates of the center of mass and total momenta which are defined by  (\ref{05}), (\ref{00005}) and satisfy noncommutative algebra (\ref{07})-(\ref{08}) with effective parameters $\tilde{\theta}$, $\tilde{\eta}$. We use notation $H_{rel}$  for hamiltonian corresponding to the relative motion  and depending on the coordinates and momenta of relative motion.  As we have shown above, the motion of the center-of-mass and the relative motion are independent in noncommutative phase space in the case when conditions (\ref{condN}), (\ref{condN2}) are satisfied. In this case $\{\tilde{{\bf P}}^{2}/(2M)+M V(\tilde{X}_1,\tilde{X}_2),H_{rel}\}=0$. So,  the equations of motion for the center-of-mass read
 \begin{eqnarray}
\dot{\tilde{X}}_1=\frac{P_{1}}{M}+M\tilde{\theta}\frac{\partial V(\tilde{X}_1,\tilde{X}_2)}{\partial \tilde{X}_2},\label{form020}\\
\dot{\tilde{X}}_2=\frac{P_{2}}{M}-M\tilde{\theta}\frac{\partial V(\tilde{X}_1,\tilde{X}_2)}{\partial \tilde{X}_1},\\
\dot{\tilde{P}}_{1}=-M\frac{\partial V(\tilde{X}_1,\tilde{X}_2)}{\partial \tilde{X}_1}+\tilde{\eta}\frac{P_2}{M},\\
\dot{\tilde{P}}_{2}=-M\frac{\partial V(\tilde{X}_1,\tilde{X}_2)}{\partial \tilde{X}_2}-\tilde{\eta}\frac{P_1}{M}.\label{form021}
\end{eqnarray}
Note, that equations of motion depend on the effective parameters of noncommutativity which according to definitions (\ref{eff}), (\ref{eff2}) depend on the composition of macroscopic body. This fact is an additional cause of violation of the weak equivalence principle in noncommutative phase space. Taking into account conditions (\ref{ondN}), (\ref{ondN2}), we can write
 \begin{eqnarray}
\dot{\tilde{X}}_1=\frac{P_{1}}{M}+\gamma\frac{\partial V(\tilde{X}_1,\tilde{X}_2)}{\partial \tilde{X}_2},\\
\dot{\tilde{X}}_2=\frac{P_{2}}{M}-\gamma\frac{\partial V(\tilde{X}_1,\tilde{X}_2)}{\partial \tilde{X}_1},\\
\dot{\tilde{P}}_{1}=-M\frac{\partial V(\tilde{X}_1,\tilde{X}_2)}{\partial \tilde{X}_1}+\alpha P_2,\\
\dot{\tilde{P}}_{2}=-M\frac{\partial V(\tilde{X}_1,\tilde{X}_2)}{\partial \tilde{X}_2}-\alpha P_1.
\end{eqnarray}
So, in the case when conditions (\ref{ondN}), (\ref{ondN2}) are satisfied the motion of a body in gravitational field does not depend on its mass and composition and the equivalence principle is recovered in noncommutative phase space.

Let us stress that besides recovering of the weak equivalence principle, conditions (\ref{ondN}), (\ref{ondN2}) give the possibility to solve at least two problems in noncommutative phase space. Namely, if the conditions are satisfied the motion of the center-of-mass of composite system and relative motion are independent, the effective parameters of noncommutativity (\ref{eff}), (\ref{eff2}), corresponding to composite system, do not depend on its composition.
In addition in the next section we show that the same conditions can be also obtained from the independence of kinetic energy of composite system on its composition and  from the additivity property of the kinetic energy.

At the end of this section it is worth noting that equivalence principle can be also recovered in deformed space with minimal length $[\hat{X},\hat{P}]=i\hbar(1+\beta\hat{P}^{2})$ with the help of relation of the parameter of deformation $\beta$ with mass. Namely the equivalence principle is not violated in the deformed space when condition $\sqrt{\beta}m=\gamma=const$ holds \cite{Tkachuk}. It is important that the same condition gives the possibility to solve a list of problems in deformed space \cite{Tkachuk,Quesne,Tkachuk1}.

\section{Kinetic energy in noncommutative phase space and the parameters of noncommutativity}

Let us examine a system of particles in noncommutative phase space and consider the case when each particle of the
system moves with the same velocity as the whole system. The problem is equivalent to the motion of a macroscopic body which is divided into $N$ parts that can be treated as a point particles.
So, we study a system of $N$ particles of masses $m_a$ with parameters of noncommutativity  $\theta_a$, $\eta_a$.  The effective parameters of noncommutativity $\tilde{\theta}$, $\tilde{\eta}$ are  defined by (\ref{eff}), (\ref{eff2}). The kinetic energy of the system is the following
\begin{eqnarray}
T=\frac{\tilde{P}_1^2}{2M}+\frac{\tilde{P}_2^2}{2M},
\end{eqnarray}
here $\tilde{P}_i^{(a)}$ are the total momenta which are given by (\ref{05}), $M=\sum_a m_a$ is the total mass of the system.

In particular case of motion of a composite system in uniform gravitational field, from  (\ref{form020})-(\ref{form021}) we have
\begin{eqnarray}
\tilde{P}_1=\tilde{A}\cos\frac{\tilde{\eta}}{M}t+\tilde{B}\sin\frac{\tilde{\eta}}{M}t,\label{form028}\\
\tilde{P}_2=-\tilde{A}\sin\frac{\tilde{\eta}}{M}t+\tilde{B}\cos\frac{\tilde{\eta}}{M}t-\frac{M^2g}{\tilde{\eta}},\label{form029}
\end{eqnarray}
with $\tilde{A}$, $\tilde{B}$ being constants which in the case of initial conditions  (\ref{form030})-(\ref{form031}) are defined as
\begin{eqnarray}
\tilde{A}=M\tilde{\upsilon}_{01},\\
\tilde{B}=M\tilde{\upsilon}_{02}+\frac{M^2g}{\tilde{\eta}}-M^2g\tilde{\theta},
\end{eqnarray}
here we use the notation $\tilde{\upsilon}_{01}$, $\tilde{\upsilon}_{02}$ for the initial velocities of the center-of-mass of the system $\tilde{X}_1(0)=\tilde{\upsilon}_{01}$, $\tilde{X}_2(0)=\tilde{\upsilon}_{02}$.
So, the kinetic energy of the system can be written as follows
\begin{eqnarray}
T=
T_0+g^2M^3\left(\frac{1}{\tilde{\eta}^2}+\frac{\tilde{\theta}^2}{2}-\frac{\tilde{\theta}}{\tilde{\eta}}\right)+M^2g\tilde{\upsilon}_{02}\left(\frac{1}{\tilde{\eta}}-\tilde{\theta}\right)+\nonumber\\+
\frac{M^2g}{\tilde{\eta}}\left(\tilde{\upsilon}_{01}\sin\frac{\tilde{\eta}}{M}t+\left(\frac{M g}{\tilde{\eta}}-M g \tilde{\theta}+\tilde{\upsilon}_{02}\right)\cos\frac{\tilde{\eta}}{M}t\right),\label{form040}
\end{eqnarray}
where
\begin{eqnarray}
T_0=\frac{M(\tilde{\upsilon}^2_{01}+\tilde{\upsilon}^2_{02})}{2}
\end{eqnarray}

Let us note here that effective parameters of noncommutativity (\ref{eff}), (\ref{eff2})  depend on the parameters of noncommutativity of individual particles and on their masses. So, the kinetic energy (\ref{form040}) depends on the composition of the system. Note, that the property of independence of kinetic energy of the composition is satisfied in the case when parameters of noncommutativity corresponding to a particle are determined by its mass as (\ref{ondN}), (\ref{ondN2}). In this case we have
\begin{eqnarray}
T=
T_0+M\left[g^2\left(\frac{1}{\alpha^2}+\frac{\gamma^2}{2}-\frac{\gamma}{\alpha}\right)+g\tilde{\upsilon}_{02}\left(\frac{1}{\alpha}-\gamma\right)+\right.\nonumber\\\left.+
\frac{ g}{\alpha}\left(\tilde{\upsilon}_{01}\sin \alpha t+\left(\frac{g}{\alpha}-g \gamma+\tilde{\upsilon}_{02}\right)\cos \alpha t\right)\right].\label{form043}
\end{eqnarray}
So, the kinetic energy of the system depends on its total mass and constants $\gamma$, $\alpha$ and does not depend on its composition.

On the other hand according to the additivity property
the kinetic energy of a system reads
\begin{eqnarray}
T=\sum_{a}T_{a}=\sum_{a}\frac{(P_1^{(a)})^2}{2m_a}+\frac{(P_2^{(a)})^2}{2m_a}=\nonumber\\
=\sum_{a}\left[T_{0a}+g^2m_a^3\left(\frac{1}{\eta_a^2}+\frac{\theta_a^2}{2}-\frac{\theta_a}{\eta_a}\right)+m_a^2g\tilde{\upsilon}_{02}\left(\frac{1}{\eta_a}-\theta_a\right)+\right.\nonumber\\\left.+
\frac{m_a^2g}{\eta_a}\left(\tilde{\upsilon}_{01}\sin\frac{\eta_a}{m_a}t+\left(\frac{m_a g}{\eta_a}-m_a g \theta_a+\tilde{\upsilon}_{02}\right)\cos\frac{\eta_a}{m_a}t\right)\right].\label{form041}
\end{eqnarray}
Here we take into account that velocity of a particle in the system is the same as the velocity of the whole system.
Note that the expressions for the kinetic energy (\ref{form040}), (\ref{form041}) are different. Let us stress that in the case when conditions (\ref{ondN}), (\ref{ondN2}) are satisfied  expression (\ref{form041}) can be rewritten as
\begin{eqnarray}
T=
T_0+\sum_{a}m_{a}\left[g^2\left(\frac{1}{\alpha^2}+\frac{\gamma^2}{2}-\frac{\gamma}{\alpha}\right)+g\tilde{\upsilon}_{02}\left(\frac{1}{\alpha}-\gamma\right)+\right.\nonumber\\\left.+
\frac{ g}{\alpha}\left(\tilde{\upsilon}_{01}\sin \alpha t+\left(\frac{g}{\alpha}-g \gamma+\tilde{\upsilon}_{02}\right)\cos \alpha t\right)\right].\label{form044}
\end{eqnarray}

 So, if parameters of noncommutativity, corresponding to a particle, are determined by its mass as (\ref{ondN}), (\ref{ondN2}),  expressions for the kinetic energy (\ref{form043}), (\ref{form044}) are the same and the kinetic energy has the additivity property.

\section{Conclusion}

In the paper we have considered the weak equivalence principle in noncommutative phase space. A general case when different particles feel noncommutativity with different parameters has been studied.
It has been shown that the motion of particle or composite system in gravitational field depends on its mass. Therefore,  the equivalence principle is violated in noncommutative phase space. We have proposed the conditions on the parameters of noncommutativity (\ref{ondN}), (\ref{ondN2}) which give the possibility to recover the equivalence principle.
Moreover, the same conditions have been derived from the additivity property of kinetic energy of composite system in noncommutative phase space and also from independence of the kinetic energy of the systems composition.

In addition, we have shown that in the case when parameters of noncommutativity corresponding to a particle are determined by its mass as (\ref{ondN}), (\ref{ondN2}) the motion of the center-of-mass of composite system in noncommutative phase space and the relative motion are independent, also, the effective parameters of noncommutativity, which describe motion of the  center-of-mass of composite system in noncommutative phase space, are independent of the composition of the system.

The importance of conditions (\ref{ondN}), (\ref{ondN2}) is stressed by the number of solved problems in noncommutative phase space.
 We have shown that only two conditions on the parameters of noncommutativity (\ref{ondN}), (\ref{ondN2}) gives the possibility to solve at least four problems in noncommutative phase space. Namely, if conditions (\ref{ondN}), (\ref{ondN2}) hold, the equivalence principle is recovered in noncommutative phase space, the motion of the center-of-mass of composite system and the relative motion are independent, the additivity property of kinetic energy of composite system is preserved, the kinetic energy of composite system is independent of the systems composition.

\section*{Acknowledgements}
This work was partly supported by Project FF-30F
(No. 0116U001539) from the Ministry of Education
and Science of Ukraine.

\end{document}